\documentclass[12pt]{article}

\usepackage{latexsym}

\font\tenmsbm=msbm10 scaled 1200 \font\sevenmsbm=msbm9
\newfam\msbmfam
\textfont\msbmfam=\tenmsbm \scriptfont\msbmfam=\sevenmsbm

\makeatletter \@addtoreset{equation}{section} \makeatother

\def\beq{\begin{equation}}
\def\eeq{\end{equation}}
\def\bea{\begin{eqnarray}}
\def\eea{\end{eqnarray}}
\def\bet{\begin{tabular}}
\def\eet{\end{tabular}}

\catcode`@=11
\def\quad@rato#1#2{{\vcenter{\vbox{
        \hrule height#2pt
        \hbox{\vrule width#2pt height#1pt \kern#1pt \vrule width#2pt}
        \hrule height#2pt} }}}
\def\quadratello{\mathchoice
\quad@rato5{.5}\quad@rato5{.5}\quad@rato{3.5}{.35}\quad@rato{2.5}{.25}
}

\begin{document}

\begin{titlepage}

\begin{flushright}
HUTP-02/A038
\end{flushright}

\vspace{1truecm}

\begin{center}

{\Large \bf Comments on the dependence between electric charge and
magnetic charge\footnote{ The paper is partially supported by the
National Natural Science Foundation of China and the Nature
Science Foundation of Zhejiang Province under Grant Nos 102011
and 102028}}

\vspace{2cm} Kang Li\footnote{kangli@zimp.zju.edu.cn}

\vspace{1cm}

{\it\small Department of Physics, Zhejiang University, Hangzhou, 310027, P.R. China\\
and\\
Jefferson Physical Laboratory, Harvard University, Cambridge, MA
02138, USA

\smallskip
  }

\vspace{1cm}

\begin{abstract}
\vspace{0.5cm}  By using the two 4-dimensional potential
formulation of electromagnetic (EM) field theory introduced in
\cite{LN1}, we found that the $SO(2)$ duality symmetric EM field
theory can be reduced to the magnetic source free case  by a
special choice of $SO(2)$ parameter,this special case we called
nature picture of the EM field theory, the reduction condition led
to a result, i.e. the electric charge and magnetic charge are no
more independent. Some comments to paper \cite{SD} are also
mentioned.
\end{abstract}

\end{center}
\vskip 0.5truecm \noindent


\vskip 0.5truecm \noindent

\end{titlepage}

\newpage
\baselineskip 6 mm

\section{Introduction}

Recently there has been a much interest in study of EM duality
symmetry\cite{Carneiro}\cite{LM}\cite{Mendez}\cite{Galvao}\cite{Donev},
because it plays a very important role in study of duality
symmetry in string theory\cite{Kir}. In recent papers
\cite{LN1}\cite{LN2}, we introduced a two 4-dimensional potential
formulation to describe the duality symmetry of classical EM field
theory,   the theory is Lorentz covariant, and it has  manifesting
$SO(2)$ duality symmetry, moreover, we don't have the non physical
singularities around magnetic charge, i.e. we do not need the
concept of Dirac string. As a matter of fact, we know that the
classical and quantum electrodynamics match the experiment very
well, and we never find the individual  monopoles (Magnetic
charges) before, it seems we don't need the theory which include
the magnetic sources, but as we know the one generator of  duality
group $SL(2,\mathcal{Z})$ corresponds to the EM duality, if there
is EM duality symmetry, magnetic sources must be
introduced\cite{LN1}.  Fortunately,  in the two 4-dimensional
potential formalism of the EM field theory, we find a light of
hope to solve this conflict. In this formulation, because the
theory invariant under the $SO(2)$ duality transformation, the
theories are equivalent with different choice of the $SO(2)$
parameter. In a special case, we call it a nature picture,  the
equivalent magnetic charge equals to zero, and the theory returns
to the usual magnetic source free electrodynamics. The condition
for this nature picture led to the electric charge and magnetic
charge related each other, i.e. they are dependent. The paper is
organized as follows. In the next section, we will give a brief
review of the two potential formulation introduced in \cite{LN1},
where Maxwell equations is written in both $SO(2)$ and Lorentz
covariant way. In third section, that is the main section of this
paper, we will explain how 2 potential formulation can be reduced
to usual one potential formulation in the nature picture, and
give out the explicit relationship between electric charge and
magnetic charge, and the general expressions of Lorentz force and
Aharonov-Bohm phase factor in nature picture as well as in general
$\theta$ picture are also discussed. Finally there are some
discussion and short comment to reference \cite{SD} are  given in
the last section.

\section{The review of two potential vector formulation}

For completeness, we will give a brief review of the two potential
formulation of EM field theory introduced in \cite{LN1}. Under
this formulation, the main results of this letter can be obtained
in the next section.

As we know, in the usual one potential formalism, the Maxwell
equations have duality symmetry in the source free case, but this
duality symmetry will broken down immediately when the electric
source switch on, in order to recover the duality symmetry, the
magnetic sources must be introduced. The magnetic charge called
magnetic monopole was first introduced by Dirac\cite{Dirac}, but
in usual one potential formulation there must exist a  line around
the magnetic charge on which the vector potential is singular.
This is the so called Dirac string. The advantage of the two
potential formulation is that the theory is $SO(2)$ duality
covariant and no need to use the non-physical concept of Dirac
string. Now let's review this theory briefly.

Besides the usual definition of 4-dimensional potential which we
called $A_\mu^1$, i.e.
\begin{equation}
A_\mu^1=(\phi_1,-\mathbf{A_1}),~{\rm or}~ A^{\mu 1}
=(\phi_1,\mathbf{A_1}),
\end{equation}
we also introduce
\begin{equation}
A_\mu^2=(\phi_2,-\mathbf{A_2}),~{\rm or}~ A^{\mu
2}=(\phi_2,\mathbf{A_2}),
\end{equation}
where $\phi_1$ and $\mathbf{A}_1$ are usual electric scalar
potential and magnetic vector potential in electrodynamics, while
the newly introduced potential $\phi_2$  is the scalar potential
associated to  the magnetic field and $\mathbf{A}_2$ is a vector
potential associated to the electric field. Using these
potentials, the electric field strength $\mathbf{E}$ and the
magnetic induction $\mathbf{B}$ are then expressed as:

\begin{equation}
\mathbf{E}=-\nabla\phi_1 -\frac{\partial\mathbf{A_1}}{\partial
t}+\nabla\times\mathbf{A_2},
\end{equation}
\begin{equation}
\mathbf{B}=\nabla\phi_2 +\frac{\partial\mathbf{A_2}}{\partial
t}+\nabla\times\mathbf{A_1}.
\end{equation}
In the magnetic source free case, $\phi_2$ and $\mathbf{A}_2$ are
expected to be zero, so the above equation returns to the usual
magnetic source free case.

By introducing two field tensors as
\begin{equation}
F_{\mu\nu}^I=\partial_\mu A_\nu^I -\partial_\nu A_\mu^I,~~~~I=1,2.
\end{equation}
and choosing Lorentz gauge $\partial^\mu A_\mu^I=0 $, then Maxwell
equations in the case of existing both electric and magnetic
sources, i.e.,
\begin{equation}
\nabla\cdot\mathbf{E}=\rho_e,~~~~~~\nabla\times\mathbf{B}=\mathbf{j_e}+\frac{\partial{\mathbf{E}}}{\partial{t}},
\end{equation}
\begin{equation}
\nabla\cdot\mathbf{B}=\rho_m,~~~~~~~~~~\nabla\times\mathbf{E}=-\mathbf{j_m}-\frac{\partial{\mathbf{B}}}{\partial{t}},
\end{equation}
can be recast as
\begin{equation}
\partial^\mu F_{\mu\nu}^I=g^{II'}J_\nu^{I'} ,
\end{equation}
where
$$g^{II'}=\left(\begin{array}{cc}1&0\\0&-1\end{array}\right).$$
and
\begin{equation}
J_\mu^1=J_\mu^e=(\rho_e, -\mathbf{j_e}),~~J_\mu^2=J_\mu^m=(\rho_m,
-\mathbf{j_m}).
\end{equation}
In this formulation the currents are manifestly conserved:
\begin{equation}
\partial^\nu J_\nu^I \propto\partial^\nu\partial^\mu
F_{\mu\nu}^I=0.
\end{equation}

Where the index $I$ can be regarded as $SO(2)$ index \cite{LN2},
so this formulation has manifestly $SO(2)$ duality symmetry and it
related to the gauge transformations $A^{I}_{\mu} \rightarrow
A^{I}_{\mu}+
\partial_{\mu}\chi^{I}$.

From (2.3, 2.4) and the definition (2.5) we can obtain,
\begin{equation}
E_i=F_{0i}^1+{^\ast} F_{0i}^2 ,
\end{equation}
and
\begin{equation}
B_i={^\ast}F_{0i}^1- F_{0i}^2 .
\end{equation}
So it is convenient to define a new field tensor as
\begin{equation}
F_{\mu\nu}=F_{\mu\nu}^1 +{^\ast} F_{\mu\nu}^2 ,
\end{equation}
\begin{equation}
\widetilde{F}_{\mu\nu}={^\ast} F_{\mu\nu}^1 - F_{\mu\nu}^2 ,
\end{equation}
where $ \widetilde{F}_{\mu\nu}$ is exactly the Hodge star dual of
$F_{\mu\nu}$. As we shall see, using these new field tensors we
can easily express the duality symmetry in a compact fashion. It
is easy to see that $F_{\mu\nu}$ is the analog to the usual
electromagnetic tensor defined in classical electrodynamics,
because they have exactly the same matrix form in terms of the
field strengths. Since the vector potentials in our formalism have
no singularities, one has $\partial^\mu \,{^\ast} F_{\mu\nu}^I=0$,
so Maxwell's equations can also be written as
\begin{equation}
\begin{array}{l}
\partial^\mu F_{\mu\nu}=\partial^\mu
F_{\mu\nu}^1=J_{\nu}^1\\
\partial^\mu \widetilde{F}_{\mu\nu}=-\partial^\mu
F_{\mu\nu}^2=J_{\nu}^2\\
\end{array}
\end{equation}

we can check that the potential functions defined above satisfy
the following equations,

\begin{equation}
\begin{array}{l}
\frac{\partial^2}{\partial t^2} \phi_1 -\nabla^2\phi_1=\rho_e ,\\ ~\\
\frac{\partial^2}{\partial
t^2}\mathbf{A_1}-\nabla^2\mathbf{A_1}=\mathbf{j_e}\\ ~~~~\\
\frac{\partial^2}{\partial
t^2}\phi_2 -\nabla^2\phi_2=-\rho_m,\\
~~~~\\
\frac{\partial^2}{\partial
t^2}\mathbf{A_2}-\nabla^2\mathbf{A_2}=-\mathbf{j_m}\\
\end{array}
\end{equation}

In the static case, i.e. when the sources do not depend on time
$t$, we can write
\begin{equation}
\rho_I=\rho_I(\mathbf{x}),~~ \mathbf{J}_I=\mathbf{J}_I
(\mathbf{x}),~~~~ I = 1~ and~ 2,
\end{equation}

where $I=1,2$ represent $I=e,m$ respectively. Then exactly as it
is done in the standard classical electrodynamics \cite{Jackson},
the solutions of equations ( 2.16) are given by
\begin{equation}
\phi_I=\frac{1}{4\pi}g^{II'}\int
\frac{\rho_{I'}(\mathbf{x}')}{r}d^3x'
\end{equation}
\begin{equation}
\mathbf{A}_I(\mathbf{x})=\frac{1}{4\pi}g^{II'}\int
\frac{\mathbf{J}_{I'}(\mathbf{x}')}{r}d^3x' ,\end{equation} where
$r=|\mathbf{x}-\mathbf{x'}|$, then from equations (2.3) and (2.4)
we find that the field strengths have the following representation

\begin{equation}
\mathbf{E}(\mathbf{x}) =\frac{1}{4\pi}\int\rho_e
(\mathbf{x}')\frac{\mathbf{r}}{r^3}\,d^3x'\\
-\frac{1}{4\pi}\int\mathbf{J}_m (\mathbf{x}')\,{\bf
\times}\,\frac{\mathbf{r}}{r^3}\,d^3x'
\end{equation}
\noindent and
\begin{equation}
\mathbf{B}(\mathbf{x}) =\frac{1}{4\pi}\int\rho_m
(\mathbf{x}')\frac{\mathbf{r}}{r^3}\,d^3x'\\
+\frac{1}{4\pi}\int\mathbf{J}_e (\mathbf{x}')\,{\bf
\times}\,\frac{\mathbf{r}}{r^3}\,d^3x'.
\end{equation}

\section{The dependence of the electric charge and magnetic charge}

This section is the main part of this letter. From the review
section above, we know that the two potential formulation of EM
field theory has manifestly $SO(2)$ duality symmetry, so for
different $SO(2)$ transformation parameter $\theta$, the theory
should be equivalent. We can see from the subsequence, special
choice of $\theta$, the two potential formulation can be recast to
magnetic source free one potential formulation. The special choice
of $\theta$ led to electric source and magnetic source related
each other.  This condition has also been discussed in reference
\cite{GM} in a quite different manner, but some of their results
are not correct,  we will  comment this shortly in the conclusion
section.

Let's begin with the $SO(2)$ duality symmetry of EM field theory.
The general $SO(2)$ dual transformation for $F_{\mu\nu},
\widetilde{F}_{\mu\nu}$, is given

\begin{equation}
\left(\begin{array}{c}F_{\mu\nu}^{'}\\
\widetilde{F}_{\mu\nu}^{'}\end{array}\right)=\left(
\begin{array}{cc}\cos\theta&\sin\theta\\-\sin\theta&\cos\theta\end{array}\right)\left(\begin{array}{c}F_{\mu\nu}
\\
\widetilde{F}_{\mu\nu}\end{array}\right)=R(\theta)\left(\begin{array}{c}F_{\mu\nu}
\\
\widetilde{F}_{\mu\nu}\end{array}\right)
\end{equation}
Where $$R(\theta)=\left(
\begin{array}{cc}\cos\theta&\sin\theta\\-\sin\theta&\cos\theta\end{array}\right)$$
is the general $SO(2)$ matrix. Under this transformation the
Maxwell equations is invariant and the energy density and Poynting
vectors of EM filed are invariant also.

As discussed in \cite{LN2}, we know that, under this dual
transformation, the physical quantities $(\mathbf{E},\mathbf{B}) $
and electromagnetic sources $J^{\mu 1},J^{\mu 2}$, i.e.
$\rho_e,\rho_m$ and $\mathbf{J}_e,\mathbf{J}_m$ etc. must be also
changed in the same manner. For example, we consider a dyon with
electric charge $q$ and magnetic charge $g$,  under the $SO(2)$
dual transformation, the charges change like

\begin{equation}
\left(\begin{array}{c}q^{'}\\
g^{'}\end{array}\right)= R(\theta) \left(\begin{array}{c}q\\
g\end{array}\right).
\end{equation}
that is,
\begin{equation}
q^{\prime}=q\cos\theta +g\sin\theta,
\end{equation}
\begin{equation}
g^{\prime}=-q\sin\theta +g\cos\theta .
\end{equation}
For arbitrary choice of $\theta$, after this transformation, the
electromagnetic field theory is equivalent to the theory before
transformation. For examples, when $\theta=\pi$, it is a identity
transformation,  when $\theta=\frac{\pi}{2}$, the transformation
reduced to the well known replacements as follows, $q\rightarrow
g,~g\rightarrow -q$ as well as
$F_{\mu\nu}\rightarrow\widetilde{F}_{\mu\nu},~
\widetilde{F}_{\mu\nu}\rightarrow -F_{\mu\nu},~~J_\mu^1\rightarrow
J_\mu^2,J_\mu^2\rightarrow -J_\mu^1~~  E\rightarrow B,~
B\rightarrow -E,~~$,  under this replacement, one equation changes
to another equation in the Maxwell equations group, but as a
equation group does not change, so the theory is equivalent. Now
give a special choice of parameter $\theta$ such that,
\begin{equation}
q\sin\theta =g\cos\theta ,
\end{equation}
i.e.
\begin{equation}
\theta = \arctan\frac{g}{q} ,
\end{equation}

Under this special case, which we call nature picture, we have,
\begin{equation}
g^{\prime} =0.
\end{equation}
That is, in the nature picture, the equivalent magnetic charge is
zero, and the two potential formulation given in the section above
is expected returning to the usual one potential case, if this is
true, we will arrive at a conclusion that the two potential
formulation is equivalent to the usual one potential formulation,
and the EM field theory with magnetic source is equivalent to a
magnetic source free EM field theory, the condition for this
equivalent is give by equation (3.5), i.e. the   electric charge
and magnetic charge are connected each other.  In the sequel of
this section, we will show that in the nature picture, the EM
field theory does return to the one potential usual magnetic
source free case.

Let me consider a dyon system again, in the nature picture,
electric charge and magnetic charge are related by equation (3.5),
i.e.
\begin{equation}
g=q\tan\theta .
\end{equation}
because, as we know, for a dyon system we have
$\rho_e(\mathbf{x},t)=q\delta
(\mathbf{x}-\mathbf{x}(t)),~\mathbf{j}_e=q\delta
(\mathbf{x}-\mathbf{x}(t))\frac{d\mathbf{x(t)}}{dt},~ J_\mu^1
=q\delta (\mathbf{x}-\mathbf{x}(t))\frac{d\mathbf{x_\mu(t)}}{dt}$,
and the $\rho_m ~\mathbf{j}_m ~J_\mu^2$ has the similar
expressions just need change $e$ to $m$ and $q$ to $g$, so the
condition (3.8) can also be written as
\begin{equation}
\begin{array}{l}
\rho_m =\rho_e\tan\theta \\
\mathbf{j_m}=\mathbf{j_e}\tan\theta
\end{array}
\end{equation}
or simple,
\begin{equation}
  J_\mu^2=J_\mu^1\tan\theta
\end{equation}
Because $(\rho_e,\rho_m), (\mathbf{j}_e, \mathbf{j}_m)$ and
$(J_\mu^1, J_\mu^2)$ transfer same as $(q,g)$, so in the nature
picture, we have
\begin{equation}
\rho_m^{\prime} =0,~~\mathbf{j}_m^{\prime}=0
\end{equation}
and
\begin{equation}
J_\mu^{2 \prime}=0.
\end{equation}
So the second potential $A_\mu^2$ will be vanished in the nature
picture, i.e.
\begin{equation}
  A_\mu^{2\prime}=-\frac{1}{4\pi}\int\frac{J_\mu^{2\prime}
(\mathbf{x'})}{r}d^3x'=0 ,
\end{equation}
and hence, in the nature picture, the theory recovers to the usual
magnetic source free case, and the Maxwell equation become
\begin{equation}
\begin{array}{ll}
\nabla\cdot \mathbf{E}^{\prime} =\rho_e^{\prime}, &
\nabla\times\mathbf{B}^{\prime} =\mathbf{j}_e^{\prime}
+\frac{\partial
\mathbf{E^{\prime}} }{\partial t},\\
\nabla\cdot \mathbf{B}^{\prime} =\rho_m^{\prime} =0,&
\nabla\times\mathbf{E}^{\prime} = -\frac{\partial
\mathbf{B^{\prime}} }{\partial t}.
\end{array}
\end{equation}
The $(\mathbf{E}^{\prime} ,\mathbf{B}^{\prime} )$ and the original
$(\mathbf{E}, \mathbf{B})$ are related by
\begin{equation}
\mathbf{E}^{\prime}=\mathbf{E}\cos\theta +\mathbf{B}\sin\theta,
\end{equation}
\begin{equation}
\mathbf{B}^{\prime}=-\mathbf{E}\sin\theta +\mathbf{B}\cos\theta .
\end{equation}
where $\mathbf{E},\mathbf{B}$ are give by the equations (2.20) and
(2.21), and $\mathbf{E}^\prime,\mathbf{B}^\prime$ have the similar
expressions as,
\begin{equation}
\mathbf{E}^\prime(\mathbf{x}) =\frac{1}{4\pi}\int\rho_e^\prime
(\mathbf{x}')\frac{\mathbf{r}}{r^3}\,d^3x'\\
-\frac{1}{4\pi}\int\mathbf{J}_m^\prime (\mathbf{x}')\,{\bf
\times}\,\frac{\mathbf{r}}{r^3}\,d^3x',
\end{equation}
\noindent and
\begin{equation}
\mathbf{B}^\prime(\mathbf{x}) =\frac{1}{4\pi}\int\rho_m^\prime
(\mathbf{x}')\frac{\mathbf{r}}{r^3}\,d^3x'\\
+\frac{1}{4\pi}\int\mathbf{J}_e^\prime (\mathbf{x}')\,{\bf
\times}\,\frac{\mathbf{r}}{r^3}\,d^3x'.
\end{equation}
Because $\rho_m^{\prime} =0, ~\mathbf{j}_m^{\prime} =0$, so in
nature picture we have
\begin{equation}
\mathbf{E}^{\prime}(\mathbf{x}) =\frac{1}{4\pi}\int\rho_e^{\prime}
(\mathbf{x}')\frac{\mathbf{r}}{r^3}\,d^3x'
\end{equation}
\begin{equation}
\mathbf{B}^{\prime}(\mathbf{x}) =
\frac{1}{4\pi}\int\mathbf{J}_e^{\prime} (\mathbf{x}')\,{\bf
\times}\,\frac{\mathbf{r}}{r^3}\,d^3x'.
\end{equation}
On the other hand, we can also get equation (3.19, 3.20) from
equations (3.15, 3.16) and (2.20, 2.21), for example,
$$
\begin{array}{ll}
\mathbf{B}^{\prime}&=-\mathbf{E}\sin\theta +\mathbf{B}\cos\theta \\
~&=\frac{1}{4\pi}\int (\rho_m (\mathbf{x}')\cos\theta
-\rho_e(\mathbf{x}')\sin\theta )\frac{\mathbf{r}}{r^3}\,d^3x'
+\frac{1}{4\pi}\int (\mathbf{J}_e
(\mathbf{x}')\cos\theta+\mathbf{J}_m(\mathbf{x}')\sin\theta
)\,{\bf
\times}\,\frac{\mathbf{r}}{r^3}\,d^3x'\\
~&=\frac{1}{4\pi}\int\mathbf{J}_e^{\prime} (\mathbf{x}')\,{\bf
\times}\,\frac{\mathbf{r}}{r^3}\,d^3x',
\end{array}
$$
of course the condition (3.9) has been used to get the above
result.  This again shows that the nature picture correspond to
the usual magnetic source free case.

The general two potential formulation, where both electric source
and magnetic source are considered, can be reduced to the nature
picture mentioned above, in this nature picture the EM field
theory coincides with the magnetic source free usual EM field
theory. The condition of the reduction (3.5) means that the real
electric charge and magnetic charge are related each other, i.e.
$g=q\tan\theta$ .  when $g=0, ~q\neq 0$, we have to choose $\theta
=\pi$,which is a identity transformation of $SO(2)$. Because in
this case the theory is already in the nature picture, i.e. the
magnetic source free case, so the reduction transformation should
be identity. When $g\neq 0, ~q=0$, we have to choose $\theta =\pi
/2$, the reduction transformation is give by the follow
replacement,$g\rightarrow q,  \mathbf{B}\rightarrow \mathbf{E},
\mathbf{E}\rightarrow -\mathbf{B}$. i.e. a pure magnetic source EM
field theory is equivalent to a pure electric source EM field
theory, the $SO(2)$ duality transformation matrix for this two
theory is given by
\begin{equation}
R(\pi/2)=\left( \begin{array}{ll} 0&1\\-1&0 \end{array} \right).
\end{equation}

Before finishing this section, I would like to give out the
general expression of general Lorentz force and AB phase factor in
the 2 potential formulation from their expression in the nature
picture.

It is well know that the general Lorentz force in nature picture (
magnetic source free case ) is give by,
\begin{equation}
\mathbf{F} =q^\prime (\mathbf{E}^\prime
+\mathbf{v}\times\mathbf{B}^\prime).
\end{equation}
Using the transformation (3.3, 3.15, 3.16), and the reduction
condition (3.5) we can get
\begin{equation}
\begin{array}{l}
\mathbf{F} =(q\cos\theta +g\sin\theta) [(\mathbf{E}\cos\theta
+\mathbf{B}\sin\theta)
+\mathbf{v}\times(-\mathbf{E}\sin\theta+\mathbf{B}\cos\theta)]\\
=q\mathbf{E}+g\mathbf{B}+\mathbf{v}\times
(q\mathbf{B}-g\mathbf{E}).
\end{array}
\end{equation}
This is exact the general Lorentz force expression in the presence
of magnetic source\cite{Felsager}.

Now let me discuss the Aharonov-Bohm(AB) effect\cite{AB}. As it is
know that the AB phase for a particle with electric charge $q'$,
in nature picture, is given by\cite{Li},
\begin{equation}
\begin{array}{l}
\Delta\phi =q'\int^x A_\mu^{1~\prime}dx^\mu\\
=q'[\int^x \mathbf{A}_m^{\prime}\cdot
d\mathbf{l}-\int^t\phi_e^\prime (t)dt],
\end{array}
\end{equation}
where
$$ A_\mu^{1~\prime}=\frac{1}{4\pi}\int\frac{J_\mu^{1~\prime}(x')}{r}d^3x'.$$
$J_\mu^{1~\prime}$ has the similar transformation as $q'$, so
using the transformation (3.3) and the reduction condition (3.5),
after a simple algebra,
  we can get the AB phase in the two potential formulation where  magnetic
source is considered, which reads,
\begin{equation}
\Delta\phi=q\int^xA_\mu^1dx^\mu -g\int^xA_\mu^2dx^\mu.
\end{equation}
This equation tells us that in two potential formulation the AB
phase factor contains two parts, one part is the contribution of
the electric charge as we know before, another part is the
contribution of the magnetic charge. In nature picture, this phase
factor can be equated to a equivalent electric charge AB phase
factor.

\section{Conclusion remarks}

Based on the $SO(2)$ duality symmetry of electromagnetic filed
theory, we introduced a concept of the nature picture to EM field
theory.  By a special choice of $SO(2)$ parameter (reduction
condition), the $SO(2)$ duality symmetric two potential
formulation reduced the nature picture, in this picture the EM
filed theory is the well know magnetic source free theory. In
other words, the EM field theory in the presence of magnetic
source can be equated to equivalent a magnetic source free EM
field theory.

Similar results of this paper have also been discussed in
reference \cite{SD} with much different method, their reduction
condition is same as ours , but I should comment that there exist
also some mistakes in this paper.  For example, in \cite{SD},
  if the Lorentz force is give by their
equation (39), than their Maxwell equations (40~43) should contain
electric source as well as magnetic source, but they didn't, and
then hence their transformation equations (46-47) are not correct.
The correct transformation equation should be our equations (3.15,
3.16)\footnote{To compare their equations to ours, one should keep
in mind that theirs
$(\mathbf{E},\mathbf{B},\mathcal{E},\mathcal{B})$ correspond to
ours $(\mathbf{E}^\prime,\mathbf{B}^\prime, \mathbf{E},
\mathbf{B})$.}.

From the discussion  above, we can conclude that in what I called
Nature Picture, the real magnetic charge and electric charge are
related by the reduction condition (3.8), and the electromagnetic
field in the presence of magnetic source can be equivalently
treated as the usual monopole free case, moreover the effect of
monopoles can be measured through the change of the equivalent
electric charge.

\paragraph{Acknowledgements.}

I would like to thank Professor Andrew Strominger for hospitality
in Physics Department of Harvard University where this paper was
written, and thanks are also given to Nancy Partridge for her kind
helps during my research visit to Harvard.

\end{document}